# Motion robust MR fingerprinting scan to image neonates with prenatal opioid exposure

**Author details:**

Dan Ma[1], Chaitra Badve[2], Jessie EP Sun[3], Siyuan Hu[1], Xiaofeng Wang[4], Yong Chen[3], Ameya Nayate[2], Michael Wien[2], Douglas Martin[2], Lynn T Singer[5], Jared C. Durieux[2], Chris Flask[3], Deanne Wilson Costello[6]

1. Biomedical Engineering, Case Western Reserve University, Cleveland, OH
2. Radiology, University Hospitals Cleveland Medical Center, Cleveland, OH
3. Radiology, Case Western Reserve University, Cleveland, OH
4. Quantitative Health Science, Cleveland Clinic, Cleveland, OH
5. Population and Quantitative Health Sciences, Case Western Reserve University, Cleveland
6. Neonatology, University Hospitals Cleveland Medical Center, Cleveland, OH

**Corresponding author:**

Dan Ma.

B110, Bolwell Building, 11100 Euclid Ave, Cleveland, OH, 44106

Email: dxm302@case.edu     Phone: 2168445935

**Acknowledgement**:

The authors would like to acknowledge funding from Siemens Healthineers; NIH grants EB026764-01, NS109439-01, NIDA R34 DA050341-01; CWRU Planning for the Healthy Early Development Study; NICHD1PL 1HD101059-01, HEAL Initiative:  Antenatal Opioid Exposure Longitudinal Study Consortium-Case Western Reserve University and The Hartwell Foundation. The content is solely the responsibility of the authors and does not necessarily represent the official views of the National Institutes of Health or its NIH HEAL Initiative.

**Running title : MRF for neonates with prenatal opioid exposure**

**Abstract:**

**Background**: A noninvasive and sensitive imaging tool is needed to assess the fast-evolving baby brain. However, using MRI to study non-sedated babies faces roadblocks, including high scan failure rates due to subjects' motion and the lack of quantitative measures for assessing potential developmental delays.

**Purpose**: This feasibility study explores whether MR Fingerprinting scans can provide motion-robust and quantitative brain tissue measurements for non-sedated infants with prenatal opioid exposure, presenting a viable alternative to clinical MR scans.

**Study Type:** Prospective

**Population**: Infants with prenatal opioid exposure. 13 infants

**Field Strength/ Sequence:** 3T, Clinical MRI and MRF scans

**Assessment:** MRF image quality was compared to pediatric MRI scans using a fully crossed, multiple reader multiple case study. The quantitative T1 and T2 values were used to assess brain tissue changes between babies younger than one month and babies between one and two months.

**Statistical Tests**: Generalized estimating equations (GEE) model was performed to test the significant difference of the T1 and T2 values from eight white matter regions of babies under one month and those are older. MRI and MRF image quality were assessed using Gwet's second order auto-correlation coefficient (AC2) with its confidence levels. We used the Cochran-Mantel-Haenszel test to assess the difference in proportions between MRF and MRI for all features and stratified by the type of features.

**Results**: In infants under one month of age, the T1 and T2 values are significantly higher ($p<0.005$) compared to those between one and two months. A multiple-reader and multiple-case study showed superior image quality ratings in anatomical features from the MRF images than the MRI images.

**Conclusions**: This study suggested that the MR Fingerprinting scans offer a motion-robust and efficient method for non-sedated infants, delivering superior image quality than clinical MRI scans and additionally providing quantitative measures to assess brain development.

**Introduction**

Infant neuroimaging is a growing field with the unique potential to identify key milestones in human brain development. There has been a rapid increase in the use of magnetic resonance imaging (MRI), a noninvasive method to better assess infant structural and functional brain development. Work from the Baby Connectome project has led the way in establishing baseline knowledge, such as creating standard clinical practices[1–3], connectivity maps[1,4,5], and infant brain atlases[6]. This foundational work is important for better understanding the early stages of child development as the brain doubles in size in the first year and an additional 15% the following year[7]. During this growth period the blueprint of future functional and structural neural networks, adult myelination patterns, and axonal growth is designed; all of which influence future patterns of behavior, cognitive abilities, and neuropsychiatric conditions[8–13]. With this groundwork, MRI is increasingly used to assess the impact of early developmental perturbations occurring in conditions such as preterm birth, neonatal brain insults, external injuries, and fetal teratogens[12–16], which correlates to developmental delays and neuropsychiatric disorders such as autism, schizophrenia, ADHD, or behavioral and executive function controls[12,17–21].

However, keeping young children still the whole time of an MRI scan can be difficult, especially since sedation cannot be used for research studies[22–25]. This presents a significant challenge to obtain useful imaging data within a practical scan time. Even with the optimized pediatric scan protocols that include many subject-specific features such as feed and swaddles, ear protection, mock scan preparations, and abbreviated scans, current practices often show motion effects in the resultant images, leading to a high scan failure rate[9,14,22,26]. One particularly vulnerable population, with noted motion difficulties that could benefit from motion corrected imaging techniques, are babies born with neonatal opioid withdrawal symptoms (NOWS).

There are ongoing national efforts to assess the impact of prenatal opioid exposure (POE) and other substances on infant development through the NIH Helping to End Addiction Long-Term (HEAL) initiative and the resultant Advancing Clinical Trials in Neonatal Opioid Withdrawal: Outcomes of Babies with Opioid Exposure (ACT NOW: OBOE) study[27,28] with a specific focus on neuroimaging[29–34]. Imaging for infants with NOWS, who can be more irritable and harder to soothe than unexposed babies, is particularly challenging. As a member of the ACT NOW: OBOE cohort, this particular patient population would benefit from imaging techniques that ensure clear and easy to read MRI, despite unavoidable motion during scans.

Another consideration beyond image quality is the struggle with longitudinal evaluation of developmental trajectory using neuroimaging. Although clinical MR images, such as T1 weighted and T2 weighted images, have much higher image resolution than diffusion and functional MR images, they only provide relative tissue contrast, which is highly variable and may introduce uncertainties when assessing longitudinal changes. The main outcomes of current developmental research based on clinical MRI have been limited to brain morphometry measures such as brain volumes and shapes. Obtaining additional quantitative information beyond morphometry from high-resolution images is highly desirable to assess brain development.

MR Fingerprinting (MRF)[35] is a fast quantitative MR scan for simultaneous quantification of multiple tissue property maps. Compared to clinical MRI, MRF provides several advantages for infant imaging studies. First, MRF quantifies multiple critical MR contrasts, including T1 and T2 relaxation times and myelin water fraction from a single fast scan[36]. These maps measure inherent tissue properties related to tissue composition, macromolecule concentration, iron accumulations, thus showing improved sensitivity and specificity in tissue characterization and disease diagnosis[37–41]. Second, the quantitative nature of the MRF maps enables a more reliable analysis of developmental changes of the brain longitudinally. There is a growing focus on better understanding brain developments through quantitative MRI methods, such as evaluating T1, T2, and Myelin water fraction (MWF) changes across life span[2,8,14,22,24,42,43]. MRF maps have also been used to analyze developmental changes in healthy children from 0 to five years old, providing developmental trajectories showing significant differences in T1, T2, and MWF through 20 months[1]. Third, compared to other quantitative MR methods, such as DESPOT and QRAPMASTER, MRF has shown up to 4 times higher scan efficiency and 1.4 times higher reproducibility in various studies[44–46]. Finally, the MRF design, including non-Cartesian sampling and dictionary-based mapping, has shown high tolerance to measurement errors and motion artifacts, allowing robust and artifacts-free measurements even under highly accelerated and motion corrupted scans[35,47,48].

In this study, we applied an optimized MRF scan for imaging of non-sedated babies with prenatal opioid exposure[49] as a feasibility study. We demonstrated that the MRF scan provided whole brain high resolution (0.8 mm$^3$) tissue maps from non-sedated infants with only 5 minutes of scan time. Multiple co-registered tissue property maps and synthetic MR images can be generated from a single MRF scan, which allowed quantitative evaluation of the brain tissue changes across ages. Finally, we performed a fully crossed multiple reader multiple case study to assess the image quality of MRF and pediatric MRI scans, demonstrating that MRF could become a viable MR scan for non-sedated baby imaging.

**Method:**

Infant recruitment:

Infants were recruited under an IRB approved protocol. Exposed infants included: infants born ≥37 weeks gestation with second or third trimester opioid exposure determined by maternal urine toxicology screen at delivery, maternal history, and/or infant urine, meconium, or umbilical cord toxicology screen. Exclusion criteria included: Infants with known chromosomal or congenital anomalies potentially affecting the central nervous system, Apgar score at 5 minutes of <5, any requirement for positive pressure ventilation in the NICU, inability to return for outpatient MRI and/or follow-up, intrauterine growth retardation <3rd percentile, and heavy alcohol use during pregnancy (averaging 8+ drinks per week). Informed consent was obtained by clinical site research staff from all parents/guardians of prospective study infants.

A total of thirteen infants were recruited in the study between April 2021 to December 2022, undergoing both MRI and MRF scans. The participants included three males and ten females, with twelve of them being newborns aged between 7 and 65 days. The remaining participant was a nine-month-old infant. A detailed breakdown of the infants' ages and genders can be found in Supplemental Table 1.

MRI protocol

All infants were scanned on a Siemens Vida 3.0T MRI scanner dedicated to human research studies. The MRI protocol, including subjects preparation and MRI scans, was a standardized protocol for a multisite neonate with prenatal opioid exposure study[24,49]. Only the 3D T2-weighted and 3D T1-weighted images of this MRI protocol are used in this study. For subject preparation, there was a separate room close to the scanner which provided a quiet, private space to perform safety checks and discuss informed parental/guardian consent. All imaging was conducted without sedation during natural sleep. Infants were fed, swaddled, and comfortably rested with parent(s) inside the prep or scanner room until subjects fell asleep. A trained MRI study coordinator and a neonatologist (**) monitored the sleeping child in the scanner suite. The details of the MRI sequences and scan times are listed in Table 1. The total scan time of the MRI scans was 28 minutes 44 seconds, without considering the waiting time between scans or repeated scans. If any of the scans were motion corrupted by quick visual inspection by the technician, the scan was repeated until the image quality was satisfactory. The total sequence acquisition time of the MRI protocol was kept at under 40 minutes.

MRF protocol:

The MRF scan was added after all the MRI scans, as an ancillary research scan. An MRF scan typically utilizes varying acquisition patterns including variable flip angles (FA) and repetition times (TR) in a continuous manner to acquire a series of 3D images. To maximally acquire image information from the non-sedated infants with NOWS, we applied a fast 3D MRF scan that was optimized based on a physics-inspired optimization algorithm[50]. The 3D MRF scan utilizes a stack-of-spiral acquisition[37,51], total scan time was 5 minutes (2.7 sec/slice), a whole-brain coverage of 250x250x112 mm$^3$ field of view, reconstruction matrix size 312x312x140, and 0.8x0.8x0.8 mm$^3$ isotropic image resolution.

After the MRF scan, generating quantitative maps from the 3D MRF data consisted of three main steps: First, similar to the prior MRF implementation[37,51], a dictionary was constructed using Bloch equations. The dictionary contains signals from a wide range of T1 and T2 combinations (T1 values range from 10 to 2000ms, T2 values range from 2 to 300ms). The dictionary was only simulated once and used for all the MRF scans. Second, MRF images were reconstructed using a low-rank iterative reconstruction[52]. Finally, multiple tissue property maps and synthetic-contrast MR images were generated from the same scan data. From a single MRF scan, four sets of images were generated detailed as follows:

First, quantitative T1 and T2 maps were simultaneously generated using dictionary matching[35]. An additional quantitative R1R2 map (a product of R1 and R2 maps) was generated because it provided a unique myelin sensitivity. Second, synthetic MR images[53] with various commonly used clinical contrasts, including T1-w, T2-w, fluid attenuated inversion recovery (FLAIR), and double inversion recovery (DIR), were reconstructed. Because the synthetic MR images were generated from already acquired T1 and T2 maps, they did not require additional scan time. Third, sub-voxel tissue fraction maps, including gray matter (GM), white matter (WM) and cerebral spinal fluid (CSF) maps were generated using the partial-volume MRF method[36]. The PV-MRF method was used to quantify the fractions of three main tissue types (WM, GM and CSF) from each voxel using a four-compartment model including WM (T1 = 2500 ms, T2 = 180 ms), GM (T1 = 1900 ms, T2 = 110 ms), and CSF (T1 = 5000 ms, T2 = 2000 ms) and fat (T1 = 1400 ms, T2 = 30 ms). Finally, a quantitative myelin water fraction map was generated using the PV-MRF method by assuming that each voxel of the white matter is composed of three components: water trapped between myelin layers (T1 = 130 ms, T2 = 20 ms), intracellular/extracelluar water (T1 = 1300 ms, T2 = 130 ms), and free water (T1 = 5000 ms, T2 = 2000 ms)[1]. The total reconstruction time was 4.3 hours on a standalone PC using MATLAB.

Because all the images were generated from the same MRF scan, they were inherently co-registered, removing the need for the image registrations typically performed after running a series of clinical MRI scans. The 3D images with isotropic resolution also enable the re-orientation of the images in axial, coronal, and sagittal views without the need of any additional processing steps.

Image Quality Assessment:

A multi-reader image quality assessment was performed on all thirteen subjects who received both MRI and MRF scans. Because there were no quantitative T1 or T2 maps available from the MRI protocol, the synthetic T1-weighted and T2-weighted images from MRF (MRF-T1w and MRF-T2w) were used in the assessment, to compare to the corresponding T1-weighted and T2-weighted images acquired from the pediatric MRI protocol (MRI-T1w and MRI-T2w).

We implemented a fully-crossed multiple reader multiple case design. Four image types (MRI-T1w, MRI-T2w, MRF-T1w, MRF-T2w) from thirteen neonates were randomized and rated by four readers: three board certified neuroradiologists with additional training in pediatric neuroradiology and one board certified pediatric radiologist with pediatric neuroradiology fellowship (\*\*, 5 years of experience; \*\*, 8 years of experience; \*\*, 7 years of experience; \*\*, 7 years of experience;). Image quality was assessed in three categories, with specific assessment items within each category, ranked on a 3-point scale: 1 = minor/no artifacts, well visualized in structure and myelination features; 2 = moderate artifacts, poorly defined structure and myelination; 3 = severe artifacts, structure and myelination barely identified or not seen at all. Figure 5 has the detailed list of the sixteen total evaluated items. Readers independently read all cases in their respective randomly allocated order in one session. The same independent researcher (\*\*) recorded all scores. The reader assessment protocol, description of each grade scale, and assessment form are included in the supplemental material.

**Statistical Analysis:**

A region of interest (ROI) analysis was conducted on 3D MRF maps to measure quantitative T1 and T2 values in multiple white matter regions for eleven subjects (two subjects were excluded from the ROI analysis due to the presence of severe motion artifacts in T1 and/or T2 maps). In specific, eight ROIs were drawn from the MRF maps in each subject in the frontal, parietal, temporal, and occipital white matter in both right and left hemispheres. Generalized Estimating Equations (GEE) procedure[54] was applied to model the T1 and T2 intensities derived from eight ROIs. An exchangeable correlation matrix was chosen to account for the similarity between the observations within a cluster. Comparisons were made between infants younger than one month (n = 6, 48 samples) and those aged between one and two months (n=4, 32 samples) using the GEE model.

The image quality ratings for each of the sixteen features were treated as ordinal measures. We created frequency distribution plots by image-type and by reader to show the overall observed distribution of each response. To assess the reliability and quality of MRI and MRF, we calculated Gwet's second order auto-correlation coefficient ($AC_2$) with its confidence levels by feature types and by readers[55]. $AC_2$ has the advantage of not relying on independence between observations and supports the ordinal type of data. We considered the proportion of *grade 1* (minor/no artifacts and well visualized in structure and myelination features) as the measure of image quality. We applied the Cochran-Mantel-Haenszel test to assess the association between the image type (MRI or MRF) and the image quality score (*grade 1* or higher) stratified by four readers. The CMH statistic was appropriate here, since the confounding variable, the reader, was controlled when accessing the association between the image quality and image type[56]. Bonferroni adjustment for multiple testing corrections was performed. All analysis was performed by using SAS, version 9.4 (SAS Institute, Cary, NC) and R software, version 4.2.0 (R Foundation for Statistical Computing, Vienna, Austria). The level of statistical significance was set at $p < 0.05$ (two-tailed).

**Results**:

MR Fingerprinting Maps and Synthetic Images

Figure 1 shows the quantitative T1, T2 and R1R2 maps that are directly generated from an MRF scan (pt1, 10 days). The R1R2 map highlights the image contrast of the short T1 and T2 components in the brain, e.g. myelin component, and suppress the long T1 and T2 components, e.g. cerebrospinal fluid (CSF). While each map defines the appropriate myelination pattern expected in a full-term neonate (posterior limb of internal capsule, lateral thalamus, and corona radiata region on coronal maps), the myelination is best highlighted on R1R2 maps.

Figure 2 shows synthetic MR images and tissue fraction maps that are generated from the same MRF scan. To demonstrate image quality and robustness of the 3D MRF scan, images from a different subject (pt2, 15 days) are shown. In specific, Figure 2a shows four clinical image contrasts, including T1 weighted, T2 weighted, FLAIR, and DIR, that are synthetized from the MRF scans. Importantly, the pediatric MRI protocol does not include FLAIR and DIR scans due to the scan time limit. Figure 2b shows quantitative WM, GM, CSF and MWF fraction maps. The image intensity of these four maps ranges from 0 to 1, which represents estimated fraction of each tissue present in each pixel. The myelinated posterior limb of internal capsule is best defined on the DIR and MWF images and not visualized on the white matter fraction maps. Because all the images from Figure 2 were generated from the same data, they are perfectly co-registered to each other and to the quantitative T1, T2 and R1R2 maps.

Success Rate

For all the baby MRI scans (N = 13) that were performed, four infants had at least one repeated MRI T1w scans (31% repetition rate, two infants had more than two repeated T1w scans) and four babies had at least one repeated MRI T2w scans (31% repetition rate) due to non-acceptable image quality. The detailed numbers of repeated scans are listed in Supplementary Table 1. Due to the time constraints of a single scan session, not all repeated scans achieved satisfactory image quality (Figure 3). There was no repeated MRF scans for all babies; however, two scans produced motion artifacts on the maps and were excluded in the quantitative ROI analysis.

Figure 3 compares the image quality of the MRI and MRF scans from two patients (pt3, 13 days, pt4, 7 days). For pt3 (Fig 3A, two MRI-T2w and one MRI-T1w scans were performed), the MRI T2w scan was repeated due to severe motion artifacts presented in the first scan. Although the repeated T2w image shows fewer motion artifacts, undesired blurring of cortical definition in bilateral frontal lobes is still clearly seen. The image from the following T1w scan was also corrupted by motion, as the periodic ghosting artifacts propagated through the whole brain, though no repeat scan was performed due to the scan time limit. In comparison, the synthetic T1 and T2 weighted images (MRF-T2w and MRF-T1w) of the same subject from the same slice location in Figure 3 were free of image artifacts. Figure 3B from pt4 is an example when none of the repeated scans provided acceptable images (two MRI-T2w and three MRI-T1w were performed). Patient motion during the scans caused severely blurred anatomical structure in MRI T2w image, with severe ghosting artifacts and blurriness in two repeated MRI T1w scans. In comparison, the synthetic MRF T2w and T1w scans are free of motion artifacts.

Image Quality Assessment

Figure 4 summarizes the distribution of ratings specified by the four radiologists. A total of 832 observations (16 features per image type, 13 patients, 4 radiologists) were made for both MRI and MRF images, for each image-type from all readers, and there was no missing data. Specifically, among pediatric MRI, 67.5% of T1w and 51.3% of T2w images, respectively, were ranked *grade*

*1* (minor/no artifacts and well visualized structure and myelination). While among MRF, 74.1% of T1w and 67.8% of T2w images were rated with *grade 1*. The Cochran-Mantel-Haenszel tests showed a significantly association between image quality scores and image types. The proportion of *grade 1* scores for MRF was significantly higher than for MRI, both for T1w images (p = 0.0008) and T2w images (p = 3.93e-15).

Figure 5 shows a comparison of the number of ratings for *grade 1* for each of the 16 features from each image type. For all features, MRF had significantly higher proportions of *grade 1* ratings than MRI. The two-level categorical tests using the Cochran-Mantel-Haenszel method show that MRF-T2w was rated significantly higher than MRI-T2w in both anatomical (p = 1.72e-12) and myelination (p = 0.003) categories. Similarly, MRF-T1w was also rated significantly higher than MRI-T1w in both anatomical (p = 0.048) and myelination (p = 0.013) categories.

The following two aspects of agreement were analyzed: Gwet's $AC_2$ agreement coefficients for the image features amongst the four readers, estimating the inter-reader agreement for each image feature, and $AC_2$ for each reader among all features, estimating the inter-feature agreement for each reader. All agreements were calculated between 0 and 1, scores closer to 1 indicate greater agreement. Figure 6A shows the Gwet's agreement coefficient (with 95% CIs) for each image type and each feature. The estimated $AC_2$ among all MRI image features were 0.73±0.02 for MRI-T1w and 0.67±0.03 for MRI-T2w, and the $AC_2$ among all MRF image features were 0.70±0.02 for MRF-T1w and 0.67±0.03 for MRF-T2w. In general, the inter-reader agreement was higher for T1w image types than for T2w image types, and the agreement was higher for MRI image features than MRF image features. Figure 6B shows the inter-feature agreement for each reader, measuring the consistency of each reader rating all features. Variation exists among the four readers' inter-feature agreements. The estimated $AC_2$ for reader A, B, C, D are 0.95±0.04, 0.92±0.05, 0.73±0.08, and 0.92±0.04 respectively. A stronger inter-feature agreement than inter-reader agreement indicates overall rating variability is more likely attributed to differences between radiologists than within a radiologists' own ratings.

Quantitative Assessment of Baby Brain using MRF

Figure 7A compares T1, T2, R1R2 and Myelin Water Fraction maps from two babies with age of 10 days and 9 months. The brain structure, size, and tissue properties are substantially different. Specifically, the comparison demonstrates differences in the maturation of brain structure as well as the progression of myelination. The neonatal maps demonstrate myelination of corona radiata, internal capsule, and dorsal brainstem, while the 9-month-old infant maps show progression of myelination more diffusely in the supratentorial and infratentorial white matter. The myelination patterns are best appreciated on the R1R2 and MWF maps. Figure 7B shows the average T1 and T2 values of white matter regions from eleven subjects. The 9-month-old subject (No. 11) demonstrated a significant decrease in T1 and T2 values in the white matter regions, which attributed to the myelination process, as observed in previous developmental studies[1,57]. The T1 and T2 values measured from eight white matter ROIs were compared in two groups: those aged under one month (n=6, 48 samples) and those aged between one and two months (n=4, 32 samples) using the generalized estimating equations, with an exchangeable correlation matrix chosen to account for the similarity between the observations within a cluster. The results showed that infants younger than one months exhibited significantly higher T1 and T2 values compared to those aged between one and two months (p = 7.6e-4 for T1 and p = 8.0e-5 for T2).

**Discussion:**

In this study, we performed a feasibility study of implementing MR Fingerprinting scans on babies with prenatal opioid exposure. We demonstrated that a five-minute MRF scan could simultaneously generate high-resolution (0.8 mm³ isotropic) and quantitative T1 and T2 maps, synthetic MR images, tissue segmentation maps, and myelin water fraction maps. All the images were inherently co-registered because they were generated from the same scan. The image quality of MRF regarding motion artifacts, structure, and myelin visualization was also rated higher in a multi-reader assessment by four pediatric neuroradiologists.

Motion Robustness

We demonstrated in multiple cases (Figures 3) and the image quality assessment study that MRF showed higher motion tolerance to MRI scans, even when scan times of the MRI scans (2:30 mins for T1w and 3:26 mins for T2w) were shorter than that of the MRF scans (5 mins). The high motion tolerance of MRF has been discussed in the initial MRF implementation when 20% of the scan was motion corrupted. The resulting MRF maps showed no sensitivity to the motion and nearly the same quality as the maps from the motion-free scan[35]. The inherent motion tolerance relies on the pattern recognition step of the MRF mapping, where the acquired signals are 'matched' to a predefined dictionary of predicted signals. Several studies further investigated the sensitivity of MRF scans to different types and timing of the motions[47,48]. In this study, we did not add any motion correction methods. Future studies will investigate the additional gain of motion tolerance by integrating motion correction techniques without adding the scan time[48,58].

Multi-parametric mapping and synthetic MRI

The five-minute MRF scan acquires around five to ten gigabyte data, which allowed us to generate multiple quantitative maps and synthetic images, and to quantify partial volume effects for tissue segmentation from a single scan. The quantitative maps, such as T1, T2, and MWF maps, will allow longitudinal tracking of baby development on a voxel-basis, providing more tissue-specific information than conventional morphometric analysis. In addition, synthetic MR images generated from the MRF maps provide additional clinical benefits, especially for baby scans when subject motion and scan time constraints limit the number of available image contrasts for diagnosis and research. Using the synthetic MR technique, a variety of image contrasts that highlight important tissues (Figure 2) can be generated retrospectively without adding scan time.

Comparison of MRI and MRF through image quality assessment

In this study, we evaluated the potential of MRF as a viable clinical imaging technique, focusing on comparing MRF and MRI image quality in terms of image artifacts and image features clarity. The ratings of four neuroradiologists showed that the image quality (proportion of *grade 1*) of MRF is significantly higher than MRI, in both anatomical and myelination categories. A single MRF scan providing significantly better image quality than MRI scans designed for pediatric population would make MRF a more scan efficient and preferred imaging scan.

When comparing the two image acquisition methods, the opinions of the four radiologists, as experts, play a crucial role in determining the success and expectations of MRF as a viable technique. The agreement studies helped to better understand where variability in readings may arise and where to center future efforts for reader confidence. As expected, there was lower inter-reader agreement for MRF features compared to MRI features, due to the novelty of MRF synthetic images to the radiologists. The variability in reader scores was evident from the larger number of *grade 1* for Readers A, B and D, in contrast to the larger number of *grade 3* seen by Reader C (Figure 4A). Reader C mentioned an "artificial-ness" to the MRF synthetic images in their session, indicating a default confidence level with standard MRI images, despite the possible presence of

17motion artifacts or blurring with the standard MRI images. The inter-feature agreement for each reader showed that Readers A, B and D had more consistent ratings for all image features than Reader C. Although the ratings for sixteen features need not be equal, high consistency is still expected because of the correlations between feature visualization and image quality. The expertise and confidence of the four radiologists raise interesting questions for how to train and acclimatize radiologists in the future with new imaging technology[59–61]; or even how artificial intelligence[62–64] might read imaging similarly to an experienced radiologist, and is beyond the scope of this paper.

Limitations

We focused this study on the feasibility of performing 3D MRF scans on non-sedated babies with prenatal opioid exposure and image quality assessment. Further improvements of the scans are still needed, including image quality for older babies, scan efficiency, and sensitivity to longitudinal changes. The sample size and the age range of the subjects are limited in this study due to the recruitment and follow-up challenges that arose during the COVID-19 pandemic. Therefore, we focused this study on the feasibility and image quality assessment, rather than quantitative assessment of the developmental trajectories, of the subjects. Our hospital is one of four sites participating in a national multi-site study for infants and young children with prenatal opioid exposure[49]. Therefore, we expect longitudinal studies with larger sample size in the future. Finally, the number of subjects who received the MRF scan was lower than the number of subjects who received MRI scans. This is due the fact that the MRF scan was added at the end of the pediatric MRI protocol for the multi-site study as an ancillary scan. In this study, we only assess the quality of the images from the subjects who received both MRI and MRF scans. In the future prospective studies, we will optimize the order of the scans to improve the sample size.

**Conclusion**:

We demonstrate in this study that 3D MRF is an effective scan for imaging challenging infants, particularly when motion effects are prominent. The 3D MRF scan successfully obtains quantitative and clinically useful images when pediatric MRI scans fail. The image quality of MRF regarding artifacts, anatomy, and myelination visualization is rated significantly higher than the MRI images.

20

**Table 1: MR Imaging and MR Fingerprinting protocols for the baby scans**

|  | Order | Scan Name | FOV (mm3) | Resolution (mm3) | Scan Time (min) |
|---|---|---|---|---|---|
| MRI Protocol | 1 | 3D T2-weighted | 256x256x120 | 1x1x1 | 2:45 |
|  | 2 | 3D T1-weighted-MPR | 220x220x120 | 1x1x1 | 3:26 |
|  | 3 | fMRI scan 1 | 220x220x108 | 3x3x3 | 8:02 |
|  | 4 | fMRI scan 2 | 220x220x108 | 3x3x3 | 0:16 |
|  | 5 | fMRI scan 3 | 220x220x108 | 3x3x3 | 7:06 |
|  | 6 | fMRI scan 4 | 220x220x108 | 3x3x3 | 0:11 |
|  | 7 | DTI scan 1 | 220x220x105 | 2.5x2.5x2.5 | 4:30 |
|  | 8 | DTI scan 2 | 220x220x105 | 2.5x2.5x2.5 | 0:20 |
|  | 9 | DIT scan 3 | 220x220x105 | 2.5x2.5x2.5 | 1:58 |
|  | 10 | DTI scan 4 | 220x220x105 | 2.5x2.5x2.5 | 0:10 |
|  |  |  |  |  | 28:44 in total |
|  |  |  |  |  |  |
| MRF Protocol | 11 | 3D MRF | 250x250x112 | 0.8x0.8x0.8 | 5:06 |
|  |  |  |  |  | 5:06 in total |

**Figures:**

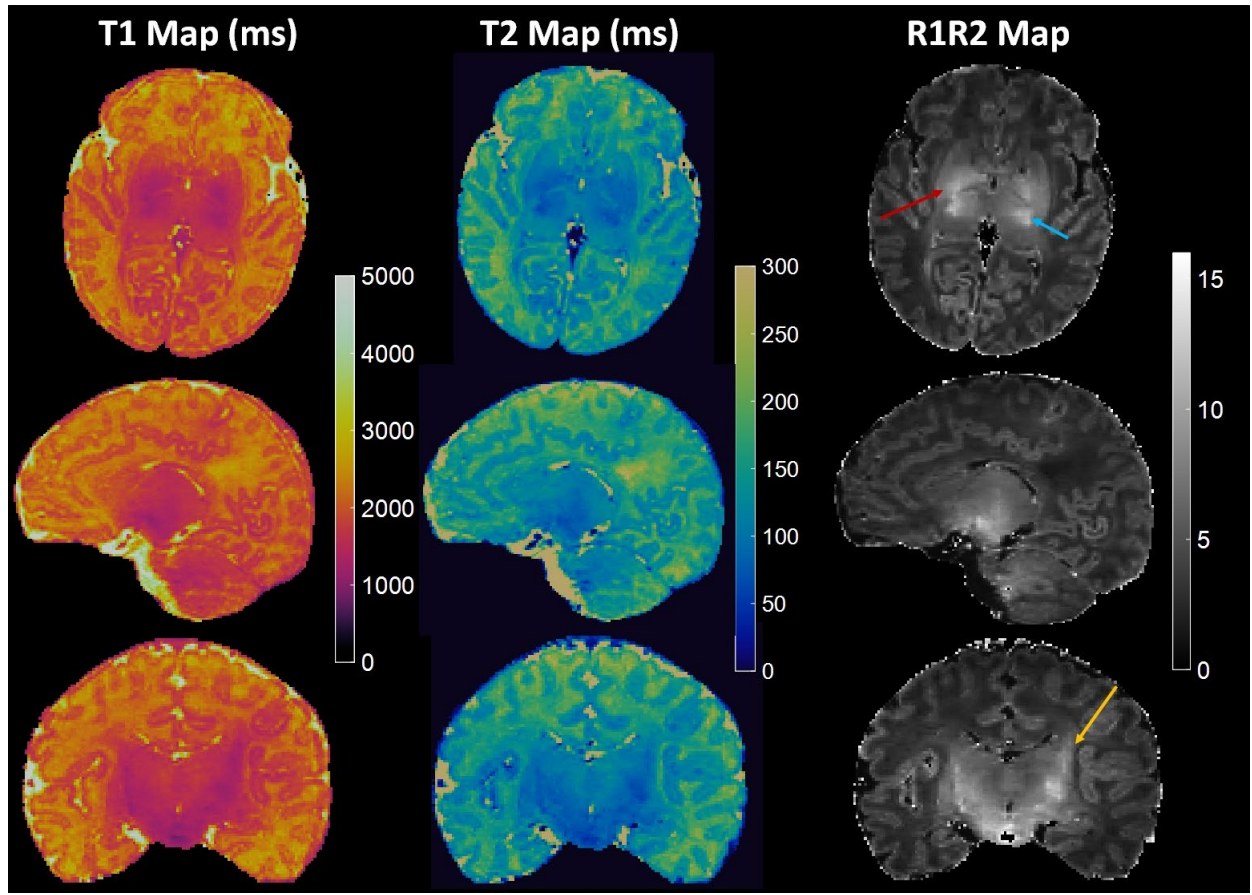

**Figure 1:** Whole brain T1, T2 and R1R2 maps generated from an MRF scan (pt1, 10 days). Each map defines the appropriate myelination pattern expected in a full-term neonate. The posterior limb of internal capsule (red arrow), lateral thalamus (blue arrow), and coronal radiata (yellow arrow) are best highlighted on R1R2 maps.

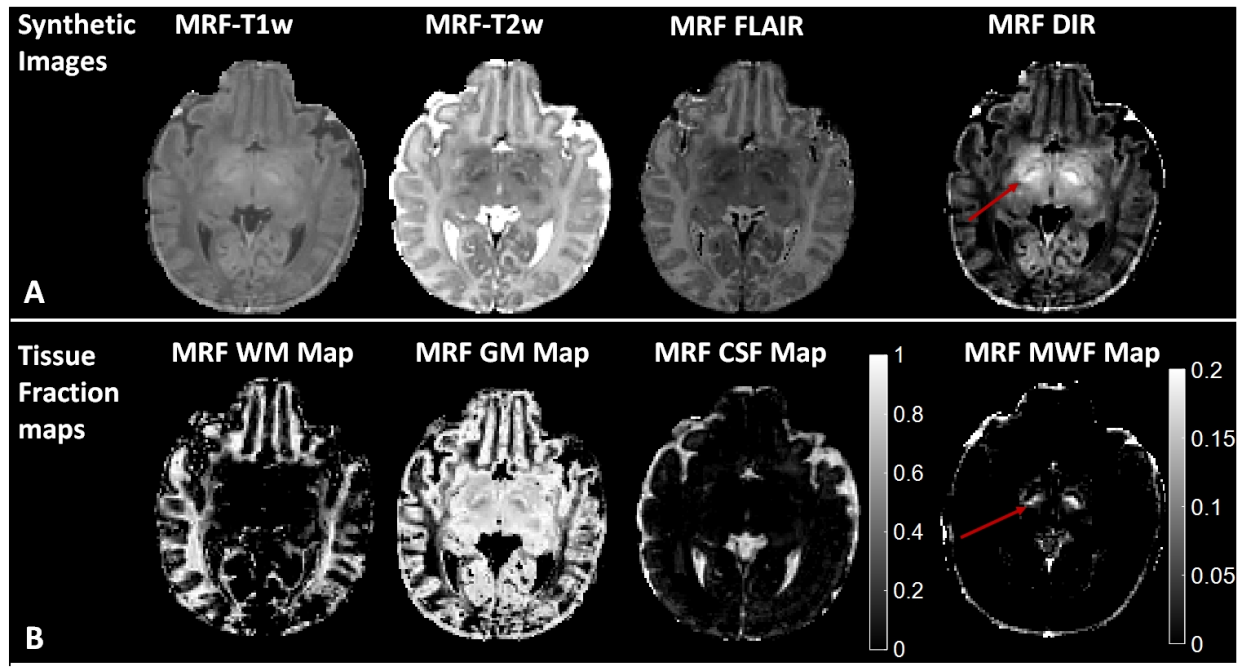

**Figure 2:** Synthetic MR images (A) and sub-voxel tissue fraction maps (B) generated from a subject (pt2, 15 days). The myelinated posterior limb of internal capsule is best defined on the DIR and MWF images (arrowed) and not visualized on the white matter fraction maps.

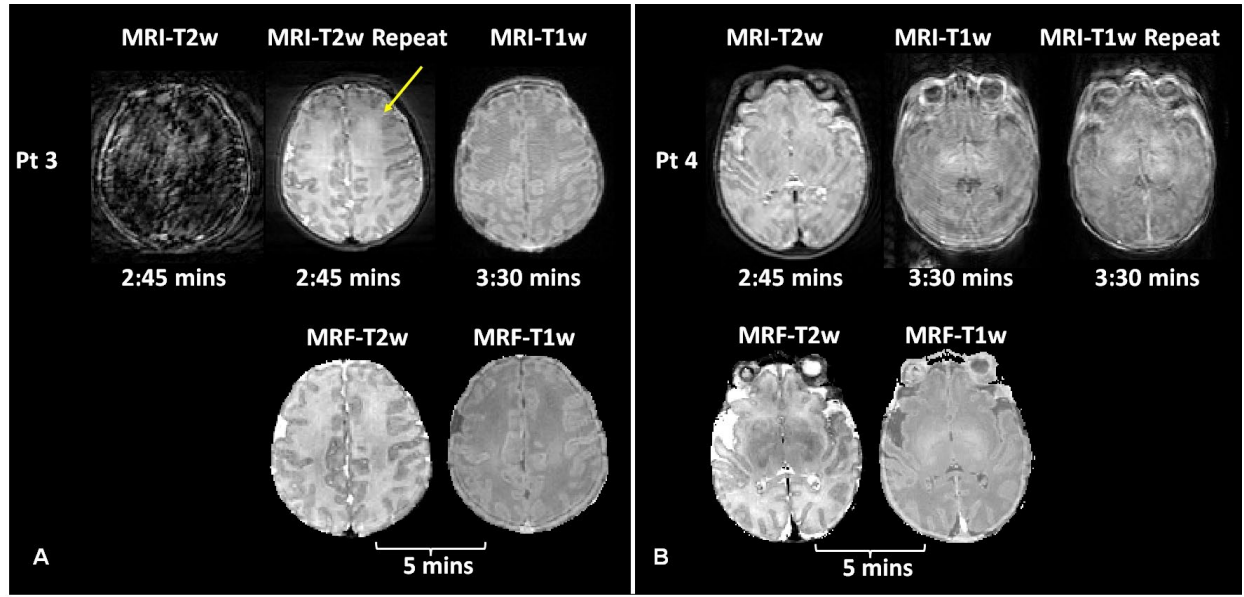

**Figure 3:** Comparisons of the image quality of the MRI-T1w and T2w images and the synthetic T1w and T2w from MRF scans from two neonate subjects. (A) (Pt3, 13 days old), The first MRI-T2w failed due to severe motion. The repeated T2w still showed shading artifacts (arrowed). The following T1w scan was also corrupted by periodic ghosting artifacts due to motion. (B) (Pt4, 6 days old), All the MRI scans (T2w and two repeated T1w) did not provide acceptable image quality. The MR images were corrupted by severe blurring and ghosting artifacts. As a comparison, the synthetic MRF T1w and T2w images from both subjects were free of motion artifacts.

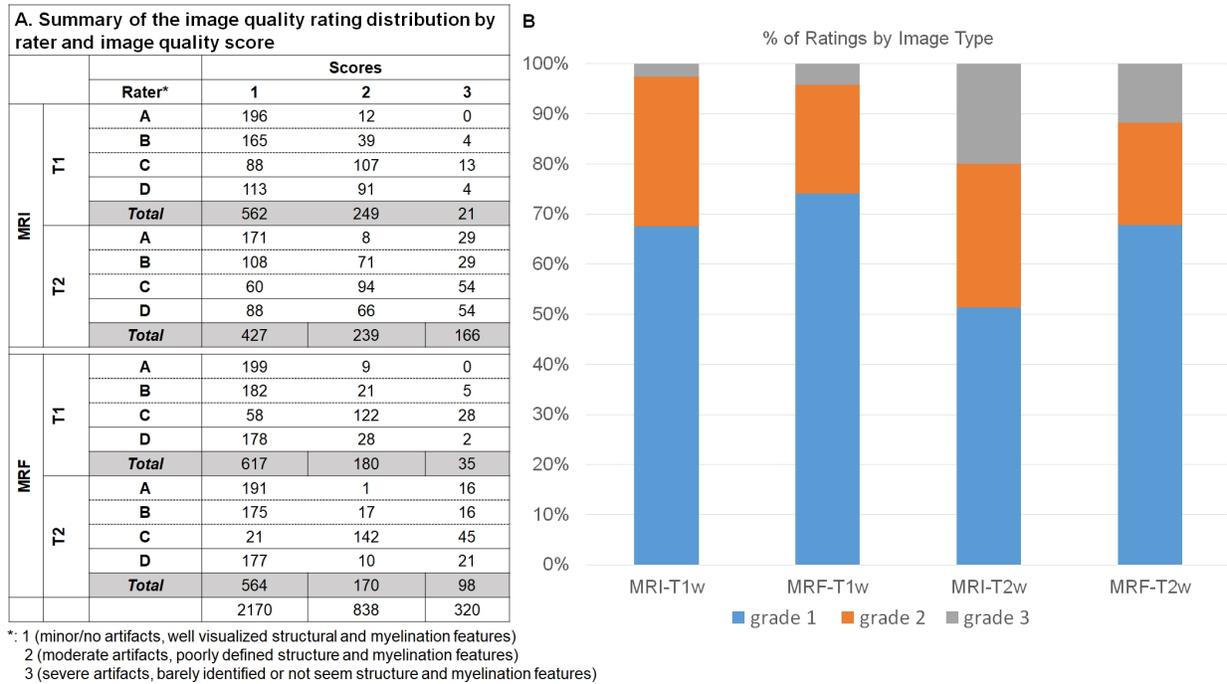

**Figure 4:** Summary of the image quality ratings for MRF and MRI by readers and image quality scores: (A) summary of the raw scores collected from the multi-reader image quality assessment by four neuroradiologists (A, B, C, D). Each number in the table is the number of times a reader rated a certain score. (B) bar plots demonstrate the distribution of the scores for each image type.

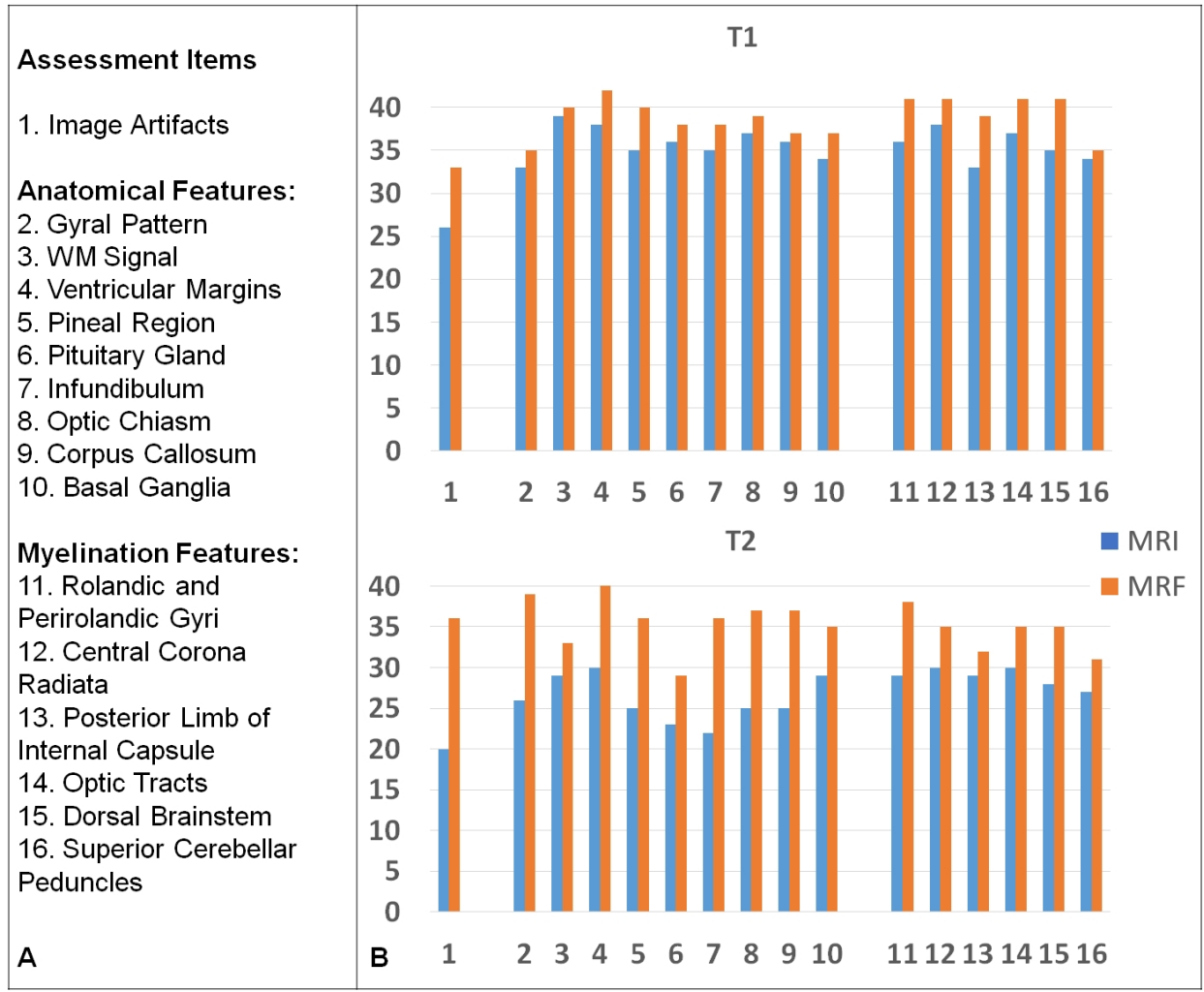

**Figure 5:** Comparison of the number of *grade 1* (minor/no artifacts and well visualized structure and myelination) ratings in each of 16 assessment image items between MRI and MRF. MRF consistently rated with lower image artifacts and better anatomical and myelin visualization compared to MRI.

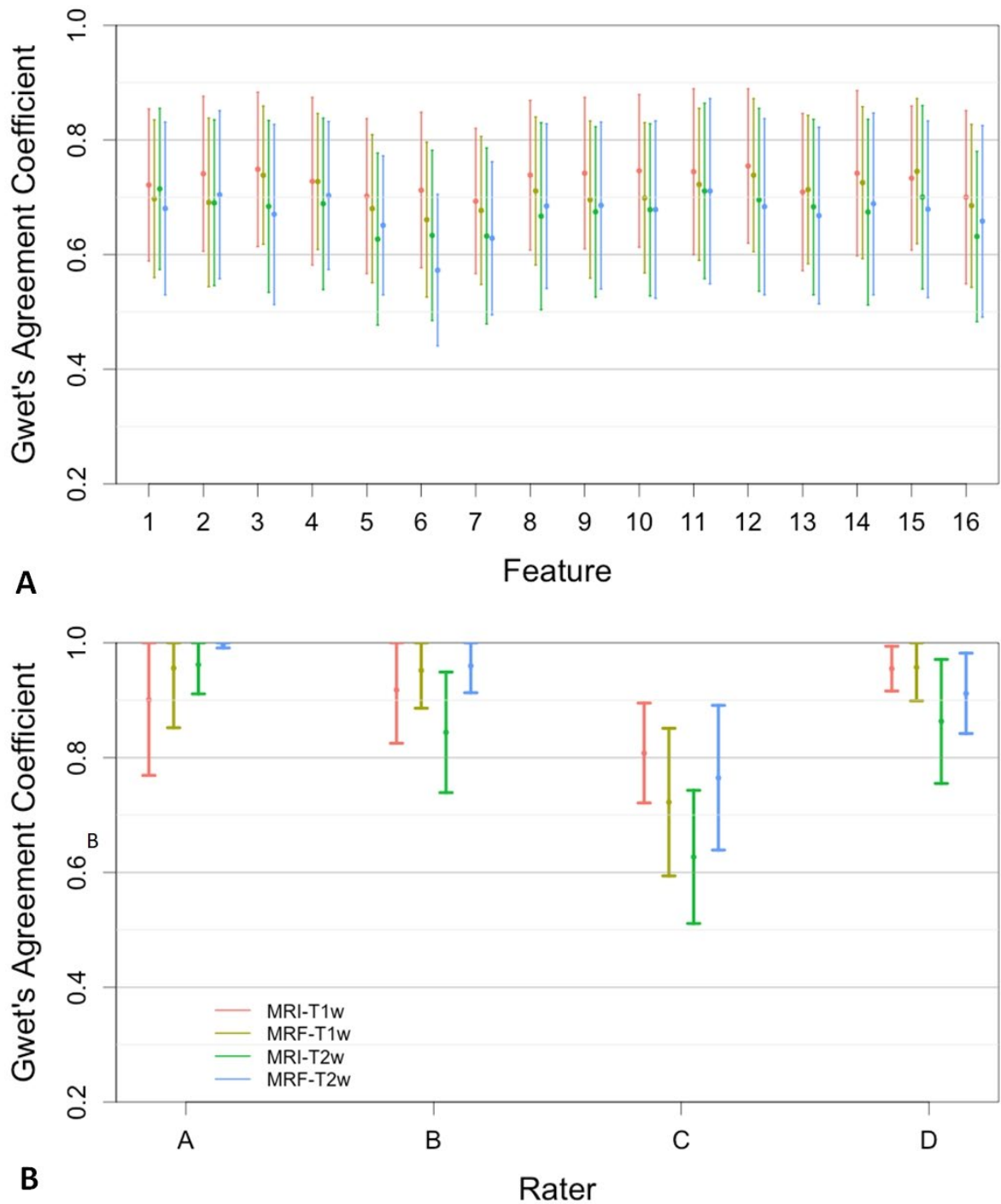

**Figure 6**: (A) Gwet's AC2 agreement coefficients for the image features amongst the four readers, estimating the inter-reader agreement for each image feature and each image type. (B) Gwet's AC2

agreement for each reader among all features in each image type, estimating the inter-feature agreement for each reader.

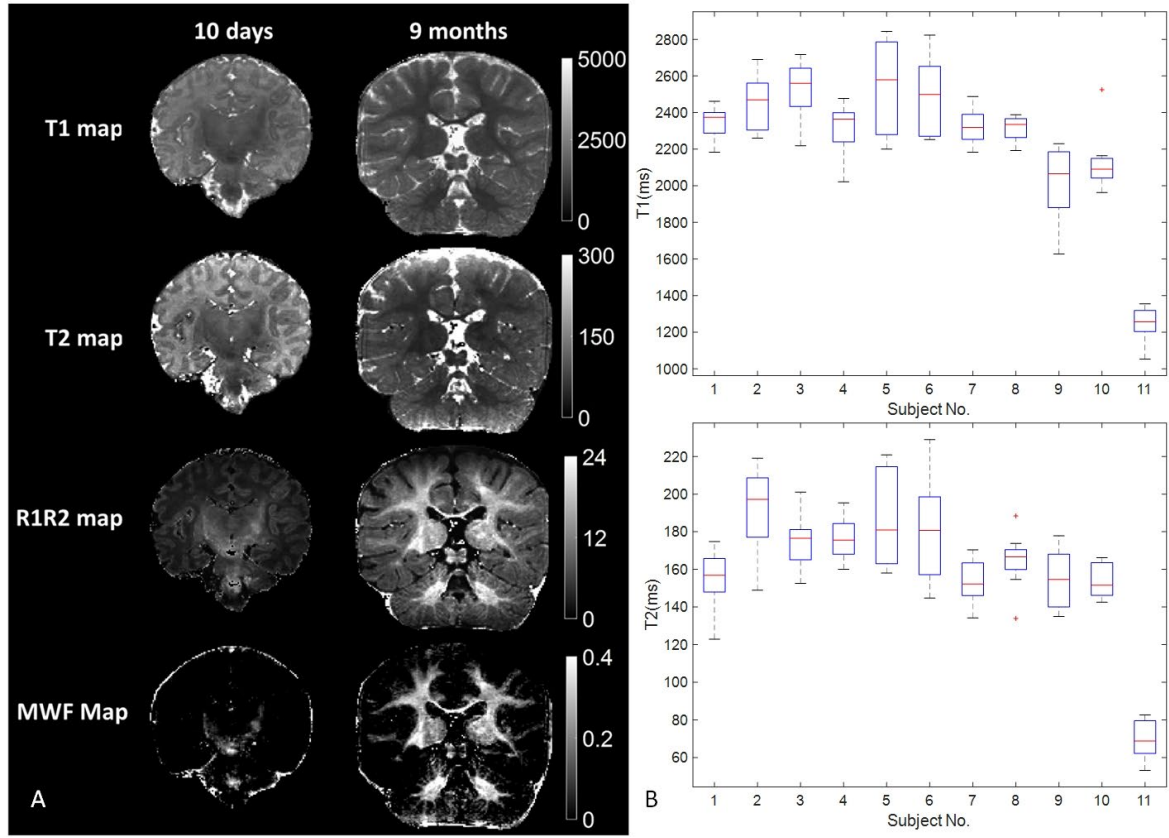

**Figure 7:** Comparison of the quantitative MRF maps from babies aged 10 days and 9 months. (A) images demonstrate differences in the maturation of brain structure as well as the progression of myelination. (B) T1 and T2 values of white matter from eleven oboe subjects. The box plot of each subject shows the T1 or T2 value distribution from eight white matter ROIs.